\documentclass[%
 reprint,
 amsmath,amssymb,
 aps,
]{revtex4-2}

\usepackage{graphicx}
\usepackage{xcolor}
\usepackage{dcolumn}
\usepackage{bm}
\usepackage{multirow}

\begin{document}

\preprint{APS/123-QED}
\title{Large exciton binding energies in MnPS$_3$ as a case study of vdW  layered magnet}

\author{Magdalena Birowska}
\email{Magdalena.Birowska@fuw.edu.pl}
 \affiliation{University of Warsaw, Faculty of Physics, 00-092 Warsaw, Pasteura 5, Poland}
 
\author{Paulo E. Faria Junior}%
\email{fariajunior.pe@gmail.com}
\affiliation{Institute for Theoretical Physics, University of Regensburg, 93040 Regensburg, Germany}%

\author{Jaroslav Fabian}%
\email{jaroslav.fabian@ur.de}
\affiliation{Institute for Theoretical Physics, University of Regensburg, 93040 Regensburg, Germany}%


\author{Jens Kunstmann}%
 \email{jens.kunstmann@chemie.tu-dresden.de}
\affiliation{Department of Chemistry and Food Chemistry TU Dresden, 01062 Dresden, Germany}%

\date{\today}

\begin{abstract}

Stable excitons in semiconductor monolayers such as transition-metal dichalcogenides (TMDCs) enable and motivate fundamental research as well as the development of room-temperature optoelectronics applications. The newly discovered layered magnetic materials present a unique opportunity to integrate optical functionalities with magnetism. We predict that a large class of antiferromagnetic semiconducting monolayers of the MPX$_3$ family exhibit giant excitonic binding energies, making them suitable platforms for magneto-optical investigations and optospintronics applications.
 Indeed, our investigations, based on first principles methods combined with an effective-model Bethe-Salpeter solver, show that excitons in bare Neel-MnPS$_3$ are bound by more than 1 eV, which is twice the excitonic energies in TMDCs. 
In addition, the antiferromagnetic ordering of monolayer samples can be inferred indirectly using different polarization of light.
\end{abstract}

\maketitle



The research on the atomically thin materials has emerged to be one of the most active topics since the discovery of graphene. 
The family of two-dimensional (2D) materials has grown rapidly and now comprises a broad class of structures including  metals, semimetals, semiconductors, insulators, and topological insulators. Only recently, in 2017, magnetic materials have been added to the family of 2D crystals, when the intrinsic ferromagnetism was experimentally reported for CrI$_3$ \cite{Huang2017} and Cr$_2$Ge$_2$Te$_6$ \cite{Gong2017}. This breakthrough triggered the interest in searching for other 2D materials with intrinsic magnetic ordering \cite{doi:10.1002/adma.201900065, Gibertini2019,Gongeaav4450,Chen983}. These 2D magnetic materials are important for fundamental physics, as well as for technological applications
in non-volatile information storage,  spin injection, filtering, and detection on atomically thin scale.

One source of 2D magnetic materials are layered magnets \cite{Gongeaav4450}, consisting of vertically stacked layers, weakly bonded via van der Waals forces. These weak forces enable the low cost fabrication of 2D magnets by using the mechanical exfoliation technique instead of the more sophisticated molecular beam epitaxy method. Furthermore, combining two different types of 2D materials in one heterostructure can result in new phenomena or new features, not present in the individual layers \cite{NatureJENS, Birowska_2019, Zutic2019MT, Zollner2019PRB, Szuplewska2019}. Similarly, combining various 2D magnets in vdW heterostructures may result in exotic spin textures and new properties not exhibited in adjacent layers leading to new types of spintronic devices.

Metal phosphorus trichalcogenides (MPX$_3$, with M = V, Cr, Mn, Fe, Co, Ni etc. and X = S, Se, Te) are vdW magnets which were widely studied in the bulk form already in the 80s and 90s of the last century \cite{LEFLEM1982455, BREC19863}.  Ultrathin layers have been obtained by using chemical vapor deposition and exfoliation methods and are the subject of the research of recent years (see the review \cite{doi:10.1002/adfm.201802151}). This class of materials exhibit intrinsic antiferromagnetism (AFM) with diverse magnetic configurations of the transition metal ions \cite{PhysRevB.46.5425}, such as MnPS$_3$ with the Neel Temperature of 78 K \cite{doi:10.1143/JPSJ.52.3919}.
  The research on MPX$_3$ is not only crucial from the point of view of fundamental studies \cite{Kang2020Nat}, but as recent reports show these materials can be a platform for field effect transistors \cite{Long2017ACS,Jenjeti2018}, photodetectors \cite{Du2016,doi:10.1002/adfm.201701342}, or tunneling devices \cite{Lee2016APL}. Important for experimental investigations, monolayers are relatively stable in air \cite{Long2017ACS,Neal2019PRB,Kim20192Dmat}.

In this study, we systematically investigate various magnetic states of bulk and monolayer MnPS$_3$ as the representative material of magnetic vdW layered MPX$_3$ systems. Starting from first-principles calculations based on the DFT+U scheme, we identify the position in k-space of the conduction (valence) band minima (maxima) and their (anisotropic) effective masses.  Furthermore, we determine material specific static dielectric constants and polarizabilities, the building blocks for the screened electron-hole interaction. The impact of the U parameter is also investigated, showing moderate influence on the optical properties. Finally, using the calculated DFT quantities combined with the versatile formalism of the effective BSE, we explore the exciton binding energies. Our calculations reveal that excitons in bare monolayer Neel-MnPS$_3$ are bound by more than 1 eV, exceeding the values in TMDCs, advancing air stable MPX$_3$ as an exciting new playground for studying optical and magnetic properties, and their interplay.

 \begin{figure*}[ht]
\centering
\includegraphics[width=0.97\textwidth]{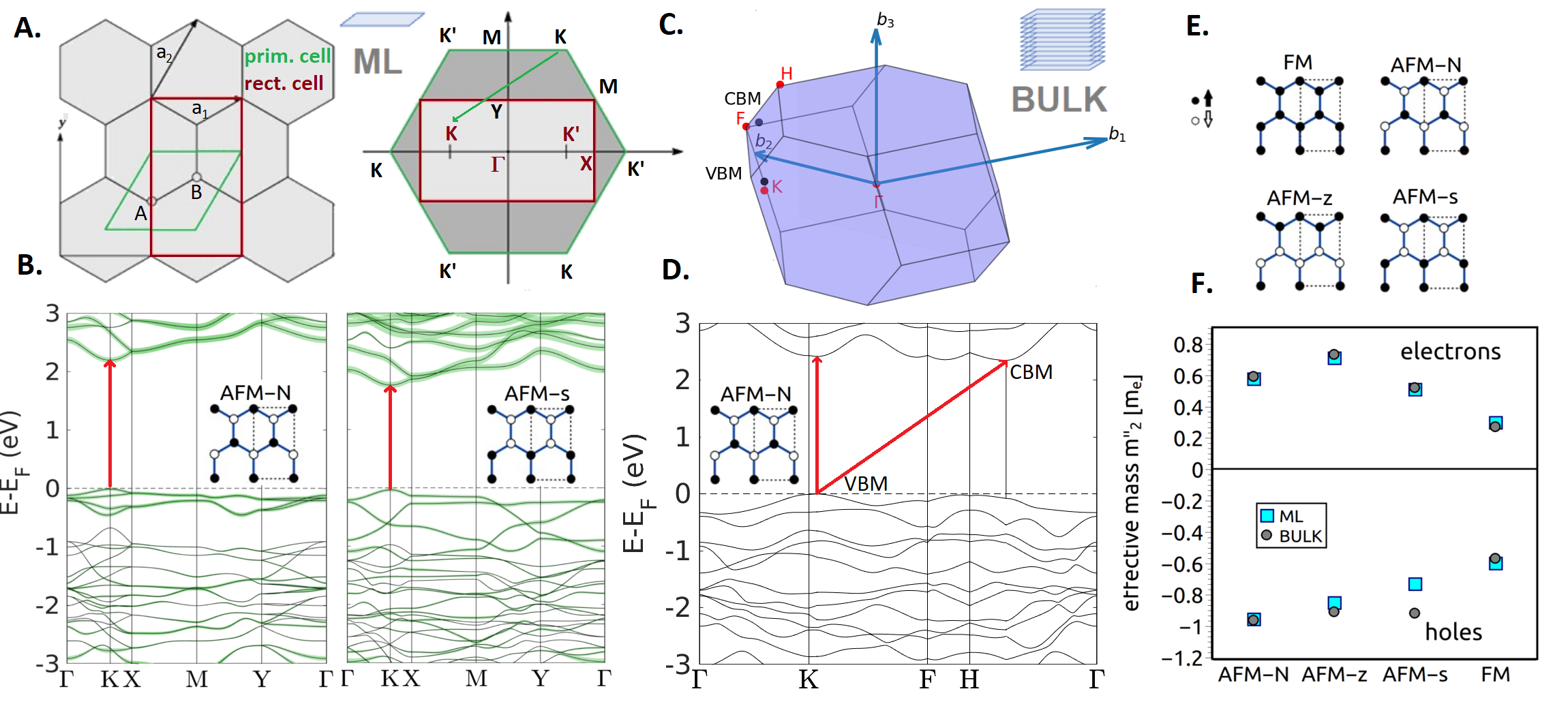}
\caption{\label{compDFT} Electronic structure of MnPS$_3$ systems. (A) Primitive (green) and extended (red) unit cells for monolayers (ML) and their corresponding first Brillouin zones (BZ) and their special points.
(B) Band structures of the magnetic states AFM-N and AFM-s (U=3eV) with projections of Mn \textit{3d} orbitals (spin up) superimposed in green. The rectangular supercell has been used to faclilitate comparison. 
(c) BZ for bulk corresponding to the primitive unit cell with special points indicated in red.
(D) Bulk band structure for the AFM-N magnetic ground state (U=5eV). VBM and CBM are located at non-special k-points which are also indicated as a black circles in (C). The red arrows indicate the band gaps. The difference between the indirect (ID, slanted arrow) and direct (D, straight arrow) band gaps is just a few tens of meV (for the exact values see SM). 
(E) Spin arrangement of the Mn atoms in the considered magnetic states. 
(F) In-plane components of the effective mass tensor of electrons and holes for U=5eV.}
\end{figure*}

\textit{DFT+U calculations}.---We consider the impact of the various magnetic states, such as the AFM-Neel (AFM-N) which is a groundstate \cite{PhysRevB.91.235425, PhysRevB.94.184428, 10.1039/C9RA09030D}, and other metastable states, i. e., AFM zigzag (AFM-z), AFM-stripy (AFM-s) and ferromagnetic state (FM) on the character of the band gap, the band alignments, and the position of the band extrema, namely valence band maximum (VBM) and conduction band minimum (CBM). In order to account for these various magnetic states of MnPS$_3$, the laterally rectangular supercell was chosen (see Fig. \ref{compDFT} (A)). Generally, our results demonstrate that the monolayers with various magnetic orderings exhibit a direct band gap at the high-symmetry point K of the first Brillouin zone (BZ), consistent with previously reported results for the ground state configuration \cite{PhysRevB.94.184428} (see Fig. \ref{compDFT} (B)). 

The differences in the band alignments, band crossing or the curvature of the bands near the VBM and CBM between the AFM-N and AFM-s magnetic configurations are clearly visible (see Fig. \ref{compDFT} (B)). Note that both the VBM and the CBM regions are composed of \textit{3d} states coming from the Mn ions. In addition, the position of the VBM and the CBM for the monolayer is affected neither by the U parameter, nor by the magnetic state (for details see Supplemental Material (SM)), where the comparison with hybrid functional HSE06 is presented). We emphasize that the DFT+U approach provides quantitatively predictive results for many correlated systems (see review \cite{doi:10.1002/qua.24521}).

Let us now focus on the bulk MnPS$_3$ calculations. The electronic band structure for the magnetic ground state (AFM-N) is presented in Fig. \ref{compDFT}(D) along high-symmetry points, with the position of the VBM and CBM indicated by the black circles in \ref{compDFT}(C).

The indirect band gaps are obtained for all magnetic states and for both U values (see Table S5). We want to emphasise, that we precisely probe the first BZ to directly state the position of the band edges. We found that the band extrema are located in non-special k-points (see SM). However, the bands close to the Fermi level are very flat (see points H and K have close energies to VBM, see Fig. \ref{compDFT} (D)). Therefore, the difference between the indirect and direct band gaps are very small, e.g. for AFM-N it is just 17 meV for U=5eV (indirect 2.352 eV vs. direct 2.369 eV). The direct transition occurs near to the “K” point (Fig. \ref{compDFT} (D)) and is in perfect agreement with previously reported  results equal to 2.37 eV (at K, with approach PBE+D2+U=5eV) \cite{10.1039/C9RA09030D}.  Due to the fact that the differences between the direct and indirect transitions are small, we introduce the term quasi-direct band gap systems for such cases. The exact coordinates of high symmetry k-points, the positions of VBM and CBM are collected in Table S5. Our calculations reveal that the position of the bands extrema are generally independent of the U parameter used. 

By properly identifying the band extrema, we can then extract the effective mass tensor for monolayer and bulk structures, for all magnetic orderings considered (see Fig. \ref{compDFT} (F)). We found that the effective masses depend strongly on the magnetic configuration. Namely, the smaller values are obtained for FM state than the masses obtained in the case of AFM configurations. For the AFM arrangements, the differences from a few tenths to a few hundredths of the rest electron mass are visible. In addition, the effective mass of holes is approximately two times heavier than the effective mass of electrons (the exact values for monolayer and bulk are given in Table S4 and Table S5, respectively). Note that the differences between the effective masses of bulk and monolayers are negligible (see Fig. \ref{compDFT} (F), where the effective mass along [010], $m_2^{\parallel}$, is compared), revealing that the mutual layered interactions of vdW-type do not alter the curvature of the valence and conduction band edges. In the case of the monolayer systems, the obtained principle reciprocal axes coincide with the Cartesian reciprocal axes, whereas in the case of the bulk systems only $m_2^{\parallel}$ is parallel to [010] crystallographic direction, the other two are shifted (see SM). In addition, the effective masses depend on the Hubbard U parameter. Generally, the smaller effective masses are obtained for U=5eV in comparison to U=3eV, which has an origin in the position of the \textit{d}-states, i. e., the U values push the \textit{d}-states away from the Fermi level (see SM). In addition, the anisotropic behaviour of the in-plane components are observed especially for AFM-s and AFM-z for monolayer as well as for the bulk systems (for the details see SM). The calculated values of the effective masses also provide important information for the transport experiments\cite{Lee2016APL}.

We now describe dielectric properties of the studied systems by means of density functional perturbation theory  in the independent particle (IP) approach neglecting local field effects \cite{PhysRevB.73.045112}. Although this approach is known to slightly overestimate the dielectric constants of semiconducting bulk crystals by up to 20 $\%$ \cite{PhysRevB.73.045112}, post-DFT schemes such as single-shot G$_0$W$_0$ approximation are still restricted to few-atoms systems and require thousands of the conduction bands to reach convergence \cite{PhysRevLett.105.146401, PhysRevB.95.035139}. Note that even fully self-consistent GW approaches \cite{PhysRevLett.99.246403,PhysRevB.95.035139,PhysRevB.98.155143} overestimates the bulk band gaps, which is likely due to underestimation of the macroscopic dielectric constant. In addition, the overestimation of the dielectric constant in the IP approach indicates that the exciton binding energy might be even larger than what is obtained in this study (in 2D systems the the exciton binding energy is inversely proportional the dielectric screening). Therefore, the IP approach is a reasonable choice for the system size considered here (up to 20 atoms in the supercell).

 Our results demonstrate that the in-plane components of static dielectric constants are isotropic for both monolayer and bulk systems. Generally, the static dielectric constants are independent of the magnetic ordering and there is only a slight dependence on the value of the Hubbard U parameter (up to 5$\%$ difference between the U=5eV and U=3eV).
Note that the layered MnPS$_3$ material has substantially  weaker screening ($\varepsilon^{\parallel}=7.75$ and $\varepsilon^{\perp}=5.88$ for U=3eV, see the Table S2) than MoS$_2$ layered system with components of dielectric constants equal to $\varepsilon^{\parallel}=15.4$ and $\varepsilon^{\perp}=7.43$ \cite{PhysRevB.84.155413}. In addition, the out-of-plane components are lower than the corresponding in-plane ones for the bulk system, a typical feature also in other layered materials. 

It is crucial to mention that the dielectric tensor is well defined for the bulk materials. In order to compare the results obtained for the bulk and monolayer systems, we calculate  the 2D polarizability $\chi_{\parallel}$ (for the details see SM). 
 Our results reveal that the polarizability does not depend on the magnetic ordering and is about 3.5  $\textrm{\AA}$ for U=3 eV (3.3 $\textrm{\AA}$ for U=5eV). 
 This value is approximately two times lower than obtained for other layered materials such as MoS$_2$ (6.60 $\textrm{\AA}$), WSe$_2$ (7.18 $\textrm{\AA}$) or MoSe$_2$ (8.23 $\textrm{\AA}$) \cite{PhysRevB.88.045318}. In addition, our results reveal negligible impact of magnetic state and the Hubbard U parameter on the dielectric properties of MnPS$_3$ systems, thus indicating that the screening is mainly performed by \textit{s, p} orbitals (U parameter affects the \textit{d} states). This is in line with the common knowledge concerning the \textit{d} orbitals, which exhibit weak screening behaviour. We have so far discussed only the electronic contribution to the dielectric screening. Since the bonding in MnPS$_3$ is partially ionic, phononic contributions are important for the total value of the static dielectric constant (see SM). However, these contributions do not seem to effectively impact the screening of the electron-hole interactions\cite{Bokdam2016}, mainly because they operate at different energy scales. We note here that the deep analysis of the phononic contribution is out of the scope of the present paper and will be studied elsewhere.

\textit{Excitonic properties}.---An exciton is the bound state formed by an electron and a hole, thus, an intrinsic  many-body phenomena. One possibility to account for the electron-hole interaction is to employ the Bethe-Salpeter equation (BSE) \cite{PhysRevLett.80.4510,Rohlfing2000PRB}. Although capable of describing realistic optical spectra, the full implementation of the BSE within first principles (using DFT energies and wavefunctions, for instance) requires a high computational cost. However, it is possible to bypass these difficulties by considering effective models, parametrized by \textit{ab initio} calculations. For instance, it has been recently shown that using the BSE approach with effective models for the band structure and for the screened electron-hole interactions, it is possible to obtain reliable results 
for excitons in monolayers (TMDCs and phosphorene, for example) \cite{Berkelbach2013PRB,ChernikovPRL2014,ChoPRB2018,Lin2020bright,FariaJunior2019PRB, Henriques2020PRBph} and 
proximitized TMDCs in vdW heterostructures\cite{PhysRevLett.119.127403, Zollner2019PRB, Zollner2020PRB, Henriques2020PRBtmdc}. Particularly in the context of 2D magnetic systems, the 
successful combination of DFT+U calculations and the effective BSE formalism in TMDC/CrI$_3$ systems by Zollner et al.\cite{Zollner2019PRB} provided reliable estimations to the
the valley Zeeman proximity exchange signatures in the optical spectra of WSe$_2$/CrI$_3$\cite{Zhong2020NN, Zhong2017SA}
and, more interestingly, to the linear dependence with respect to out-of-plane electric fields in MoSe$_2$/CrBr$_3$\cite{CiorciaroPRL2020}.

\begin{figure}[h]
\centering
\includegraphics{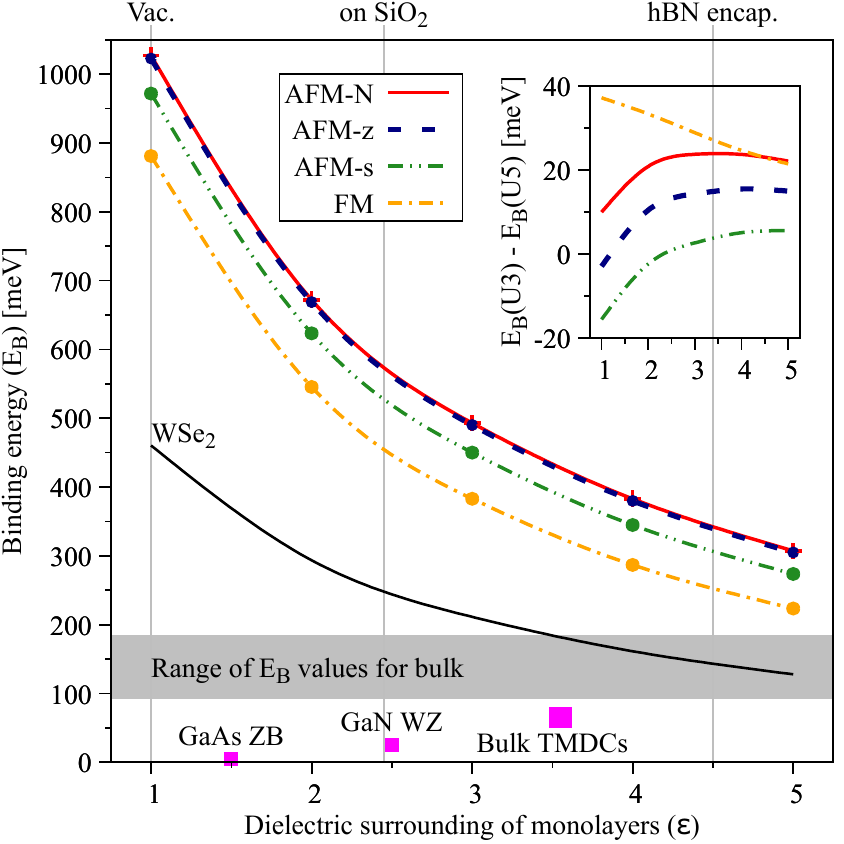}
\caption{Large exciton binding energies E$_\text{B}$ (below the band gap), for the different AFM and FM phases in monolayer and bulk MnPS$_3$. For the monolayer we show the evolution of E$_\text{B}$ as a function ofthe dielectric screening of the surroundings. The values for air/vacuum, SiO$_2$\cite{Berkelbach2013PRB} substrate and hBN encapsulation \cite{Stier2018PRL} are indicated by vertical lines. The inset on the top right corner shows the difference of E$_\text{B}$ for U=3eV and U=5eV. As a representative example of TMDC monolayers, 
we show E$_\text{B}$ for WSe$_2$ (taken from Ref~.\onlinecite{Lin2020bright}). 
The range of calculated values of E$_\text{B}$ for bulk MnPS$_3$ is shown by the shaded region. Values for typical bulk compounds are also shown for comparison: E$_\text{B}=4.2$ meV for GaAs zinc-blende \cite{AmoPRB2006}, E$_\text{B}=25.2$ meV for GaN wurtzite \cite{RodinaPRB2001} and for bulk TMDCs ranging from 50 to 
80 meV \cite{BealJPCSSP1972,GotoJPCM2000,SaigalAPL2016,Arora2015Nanoscale}. For the bulk E$_\text{B}$, the x-axis is meaningless.}
\label{fig:exciton}
\end{figure}

To investigate the band-edge excitons in bulk and monolayer MnPS$_3$ we then use the versatile formalism of the BSE with effective models   \cite{Rohlfing2000PRB,Tedeschi2019PRB,FariaJunior2019PRB}, which allows for exciton calculations with general band structure dispersion and electrostatic potentials, either for direct or indirect excitons. Particularly, we treated the conduction and valence bands with anistotropic quadratic dispersion, i.e., different effective masses along the different axes (see SM). The electron-hole interaction for monolayers is mediated by the Rytova-Keldysh potential \cite{Rytova1967MUPB,Keldysh1979JETP}, which requires the screening length of the 2D material and the effective dielectric constant, $\varepsilon$, of the surroundings to simulate effects of the substrate and/or encapsulation. For the bulk case, electrons and holes interact via the anisotropic Coulomb potential \cite{Landau2013} which requires the static dielectric constant along the different directions of the crystal. Therefore, in summary, in the effective BSE description the excitonic properties are determined by two parameters: the effective masses in conduction and valence bands, and the static dielectric function (in bulk) or the screening length (in monolayer). These properties are discussed in the DFT+U section. For the explicit form of the BSE, band dispersion, electron-hole potentials, parameters and computational details, see Sec. I B. in the SM.

In Fig.~\ref{fig:exciton} we summarize our findings for the exciton binding energies (E$_\text{B}$). Our calculations revealed that E$_\text{B}$ in bare ($\varepsilon=1$) monolayer MnPS$_3$ is quite large, reaching values slightly above 1 eV for the AFM-N phase. The other phases also show sizeable values: the E$_\text{B}$ of AFM-z has nearly the same value as the E$_\text{B}$ of AFM-N while the E$_\text{B}$ of AFM-s (FM) phase is approximately 50 (150) meV smaller than the E$_\text{B}$ AFM-N phase. These energy differences between the different phases remain quite similar in the range of effective dielectric constant of the environment. Furthermore, changing U from 3eV to 5eV preserves the same energetic ordering of E$_\text{B}$ for the different phases (the effect of U on the E$_\text{B}$ is around tens of meV, shown in the inset of Fig.~\ref{fig:exciton}).   they can be These values are around twice the values of the binding energies in conventional TMDCs (such as WSe$_2$, as shown in Fig.~\ref{fig:exciton}). The values of E$_\text{B}$ for the bulk case are also shown in Fig.~\ref{fig:exciton} to compare with the values of the monolayers, even though the x-axis (dielectric constant of the surroundings) has no meaning. We found E$_\text{B}$ in the range of 90 to 180 meV considering the different phases and values of U=3eV and U=5eV. All the calculated values for E$_\text{B}$ in the monolayer and bulk cases are given in the SM.

\begin{table}[ht]
\def\arraystretch{1.2}
\caption{Calculated optical selection rules from VBM to CBM for the 
different AFM magnetization of the monolayer systems, 
highlighting the polarization and intensity. The intensity is given by $\left|\frac{\hbar}{m_{e}}\hat{e}\cdot\vec{p}_{cv}\right|^{2}$ in which  $\hat{e}=\{\hat{x},\hat{y},\hat{z}\}$ is the light polarization and $\vec{p}_{cv}$ is the matrix element between CBM and VBM.}
\begin{tabular}{cccc}
\hline
\hline
Magnetization & U [eV] & Polarization & Intensity [$(\textrm{eV\AA})^{2}$]\tabularnewline
\hline
\multirow{2}{*}{AFM-N} & 3 & z & 0.303\tabularnewline
 & 5 & z & 0.281\tabularnewline
\multirow{2}{*}{AFM-z} & 3 & y & 0.159\tabularnewline
 & 5 & y & 0.082\tabularnewline
\multirow{2}{*}{AFM-s} & 3 & y & 0.034\tabularnewline
 & 5 & y & 0.025\tabularnewline
 \hline
\label{tab:pol}
\end{tabular}
\end{table}

For monolayer systems  the selection rules have been determined (see Table \ref{tab:pol}) for the direct transitions between VBM and CBM, following Ref. \cite{PhysRevB.101.235408}. We found that the dipole matrix elements
exhibit different non-zero components for the different magnetic states. 
Namely, the AFM-N case shows out-of-plane component (along $z$),
whereas for AFM-z and AFM-s only the in-plane component (along $y$)
is non-zero, with different intensities for each case. For the FM case, 
no optical transitions are allowed. Combined with the calculated 
exciton binding energies, the selection rules provide additional support 
that the different magnetic configurations can be distinguished optically.


\textit{Conclusions}.--- All magnetic states of monolayers employed in this study exhibit a direct band gap at the high-symmetry K points. For the bulk systems, the band gap is quasi-direct located at non-special k-points and with a moderate impact of the magnetic state on the position of the VBM and CBM. We found a strong impact of the magnetic state on the electron and hole effective masses, with moderate influence from value of the Hubbard U parameter. Generally, the in-plane components of the effective mass tensor of bulk systems and metastable anti-ferromagnetic cases of the monolayers are anisotropic. Based on the DFT+U results, we calculated the dielectric constants and 2D polarizabilities, the building blocks to model screened electron-hole interactions. The key result of our study is that the exciton binding energies are large in comparison to many other layered semiconductors such as TMDs, for both monolayer and bulk systems. In particular, we found exciton binding energies of about 1 eV for the bare monolayer, which are about two times larger than the typical values in TMDCs.  In addition, the exciton binding energies in bulk systems are about one order of magnitude larger than the ones of conventional bulk semiconductors. Analyzing the effect of the Hubbard U parameter, we found that the exciton binding energies vary on a scale of tens of meV for U=3eV to 5eV. Our results suggest that the magnetic state might be inferred indirectly from the absorption onset in optical measurements, 
and, particularly, that the antiferromagnetic configuration of monolayer systems is sensitive to the polarization of the light.

M.B. acknowledges financial support from the National Science Centre, Poland on
the basis of the decision no. DEC-2016/23/D/ST3/03446, within the framework of the research project ‘SONATA 12’ no. UMO-2016/23/D/ST3/03446. Financial support from the Deutsche Forschungsgemeinschaft (DFG, German Research Foundation) under Project-ID  314695032 (SFB 1277) is acknowledged by P.E.F.J. and J.F. and under Project-ID 317551441 by J.K. Access to computing facilities of TU Dresden (ZIH) within the project "TransPheMat", PL-Grid Polish Infrastructure for Supporting Computational Science in the European Research Space, and of the Interdisciplinary Center of Modeling (ICM), University of Warsaw are gratefully acknowledged.

\pagebreak

\begin{center}
  \textbf{\large Supplementary Material:\\Large exciton binding energies in MnPS$_3$ as a case study of vdW layered magnet}\\[.2cm]
  Magdalena Birowska,$^{1,*}$ Paulo E. Faria Junior,$^{2}$ Jaroslav Fabian,$^{2}$ and Jens Kunstmann$^3$\\[.1cm]
  {\itshape ${}^1$University of Warsaw, Faculty of Physics,\\ 00-092 Warsaw, Pasteura 5, Poland\\
  ${}^2$Institute for Theoretical Physics, University of Regensburg, 93040 Regensburg, Germany\\
  ${}^3$Department of Chemistry and Food Chemistry TU Dresden, 01062 Dresden, Germany\\}
  ${}^*$Electronic address: Magdalena.Birowska@fuw.edu.pl\\
(Dated: \today)\\[1cm]
\end{center}
\twocolumngrid

\setcounter{equation}{0}
\setcounter{figure}{0}
\setcounter{table}{0}
\setcounter{page}{1}
\renewcommand{\theequation}{S\arabic{equation}}
\renewcommand{\thefigure}{S\arabic{figure}}
\renewcommand{\thetable}{S\arabic{table}}

\tableofcontents

\section{Computational Details}
\subsection{DFT calculations}
 The bulk MnPS$_3$ compound exhibits the monoclinic space group symmetry $C2/m$. We have chosen the smallest possible supercell which reflects the particular magnetic configuration.  Namely, AFM-N and FM arrangements can be represented in primitive unit cell (10 atoms in the supercell, denoted in Table \ref{tab:Kpoints} as "p"), whereas the AFM-z and AFM-s extended (laterally rectangular supercell is used, consists of 20 atoms, denoted in Table \ref{tab:Kpoints} as "e"). The basis vectors and corresponding reciprocal lattice vectors are collected in Table \ref{tab:Kpoints}.
\begin{table*}[]
\small
\caption{\label{tab:Kpoints}The BZ is spanned by three reciprocal lattice vectors
$\vec{b_1}$, $\vec{b_2}$ and $\vec{b_3}$ constructed from the basis vectors $\vec{a_1}$, $\vec{a_2}$ and $\vec{a_3}$. The coordinates of the high symmetry "K" point for particular cell is listed in the last column. In the case of monolayer $c\to \infty$, then $p\to 0$.}  
 \def\arraystretch{1.3}
\begin{center}
\begin{tabular}{  c|  c|  c|c |c|c| c|c}
   \hline
 supercell &$\vec{a_1}$ & $\vec{a_2}$ &  $\vec{a_3}$&  $\vec{b_1}$ & $\vec{b_2}$ &  $\vec{b_3}$& K  in RLV\\
 \hline
ML "p" &(a/2 -$\sqrt{3}$a/2 0)&(a/2 $\sqrt{3}$a/2 0)&(0 0 c) & (k -m 0) & (k m 0) & (0 0 p) & (1/3 1/3 0)\\
 ML "e"  & (a 0 0)&(0 $\sqrt{3}$a 0)&(0 0 c)& (k 0 0) & (0 m 0) & (0 0 p) & (1/3 0 0) \\
Bulk "p" &(a/2 $\sqrt{3}$a/2 0)&(-a/2 -$\sqrt{3}$a/2 0)& (-x 0 $\sqrt{c^2-x^2}$) & (k m t) & (-k m -t) & (0 0 p1) & (-$\frac{1}{3}$ $\frac{1}{3}$ $\frac{2t}{3p1})$ \\  
 Bulk "e" &(a 0 0)&(0 $\sqrt{3}$a 0)&(-x 0 $\sqrt{c^2-x^2}$) & (k 0 t) & (0 m 0) & (0 0 p1) & (-1/3 0 t/(3p1)) \\
     \hline
\end{tabular}
\end{center}
\end{table*}

The calculations are performed in the framework of spin-polarised density functional theory (DFT) as implemented in VASP package \cite{PhysRevB.47.558, KRESSE199615}. The Perdew–Burke–Ernzerhof (PBE) exchange–correlation functional is  employed. The electron-ion interaction is modeled using projector augmented wave (PAW) pseudopotentials. The kinetic energy cutoff for the plane-wave expansion of the pseudo-wave function is set to 400 eV. A k-mesh of 10$\times$10$\times$2 for ML "p", 10$\times$6$\times$2 for ML "e" (10$\times$10$\times$9 for bulk "p", 10$\times$6$\times$9 for bulk "e") are taken to sample the first Brillouin zone (BZ) on $\Gamma$-centered symmetry reduced Monkhorst-Pack meshes using a Gaussian smearing with $\sigma$=0.05 eV. However, in the case of the density of states (DOS) and frequency dependent function $\varepsilon(\omega)$ the tetrahedron method was employed along with the denser k-point grids for laterally rectangular supercell  $20\times 12\times 1$ (ML "e" supercell) and $20\times 12\times 18$ (bulk "e" supercell), which was checked to be sufficient in convergence tests. 

It is well known that standard exchange correlation functionals such as LDA and GGA are insufficient to describe a non-local nature of dispersive forces, which are crucial in proper description of many systems such as layered materials \cite{PhysRevMaterials.2.034005,APP2011.Birowska} and adsorption molecules on the surfaces \cite{PhysRevLett.112.106101,BIRcomp,Acta2009,AIPMilowska}. Here, we adopted a semi-empirical Grimme method \cite{Grimme} with a D3 parametrization (DFT-D3) \cite{DFT-D3}. Note that the Grimme approach is a commonly used technique for many vdW layered materials such as TMDs \cite{Pandey2020} MPX3 \cite{PhysRevB.94.184428}. In particular, The DFT-D approach reproduce a good interlayer spacings in bulk  graphite and other 2D layered systems \cite{JCC2012R,CondMatBjorkman}. In order to model the monolayer, a 25$\textrm{\AA}$ of vacuum is added to avoid spurious interactions between replicates. The lattice parameters are fully optimized for given magnetic state. All of the atoms are relaxed until the maximal force per atom was less than $10^{-3}$ eV/$\textrm{\AA}$, and the maximal component of stress tensor is less than 0.3 kbarr for monolayers and 0.05 kbarr for bulk structures. A collinear arrangement of spins are assumed without inclusion of spin-orbit interaction.

DFT+U formalism  proposed by Dudarev \cite{PhysRevB.57.1505} is employed to properly characterize on-site Coulomb repulsion between $3d$ electrons of Mn ions, by using effective Hubbard U (U$_{eff}$ =U-J, where J=1eV). 
For each of the U, we have  fully optimized lattice parameters.

\subsubsection*{Effective masses}

In order to account for the anisotropic properties of the studied systems, we calculate the effective mass ($m^*$) tensor defined as:
\begin{equation}
\left ( \frac{1}{m^*} \right )_{ij}=\frac{1}{\hbar^2}\frac{\partial^2 E_n(\vec{k}) }{\partial k_ik_j},\:   i,j = x,y,z
\end{equation}
where $E_n(\vec{k})$ is a dispersion relation for the n-th electronic band. The second derivatives are computed numerically using finite difference method, on a five-point stencil grid as implemented in the code \cite{github}. Then, the effective mass tensor is diagonalized, and hence its components: $m_1^*$, $m_2^*$, $m_3^*$ are determined along three principle axes in reciprocal space.

\subsubsection*{Dielectric screening}
The static dielectric constant is a sum of two contributions $\varepsilon_0=\varepsilon_{\infty }+\varepsilon^p$, where $\varepsilon_{\infty }$ is electronic contribution (dielectric constant at optical frequency), where $\varepsilon^p$ is ionic response to static electric fields (phononic part). The electronic term is calculated using density functional perturbation theory  in the Independent Particle (IP) approach  neglecting the local field effects $\varepsilon^{LR}_{IP}$ \cite{PhysRevB.73.045112}, on the top of the PBE+U level. The ionic part is computed in the finite difference approach. The electronic and phononic contributions  are presented in Table \ref{tab:dielectric} and Table \ref{tab:ionic}, respectively. Note, that both contributions are of the same order, indicating that the MnPS$_3$ is partially ionic system. 
\begin{table}[h]\footnotesize
\caption{Static dielectric constants $\varepsilon_{ij}$ (electronic contribution) and 2D polarizability $\chi^{\parallel}$ calculated for the monolayer and bulk systems. The static dielectric properties are calculated by means of density functional perturbation theory  in the independent particle approach (IP) neglecting local field effects.} \label{tab:dielectric}
 \def\arraystretch{1.5}
\begin{center}
\begin{tabular}{  c|  c|  c|  c| c }
   \hline
Magn. & U [eV] &  $\chi^{\parallel}_{ML}$  [\textrm{\AA}]& $\chi^{\parallel}_{BULK}$ [\textrm{\AA}]&BULK $\varepsilon_{xx}=\varepsilon_{yy}$, $\varepsilon_{zz}$  \\
   \hline
AFM-N  & 3 & 3.45& 3.50 &  7.75  5.88   \\
 & 5 & 3.25& 3.29  & 7.33 5.67  \\
  \hline
 AFM-z & 3 &3.49& 3.49 & 7.72 5.84 \\
 & 5 &3.25& 3.44&7.60 5.82    \\
    \hline
AFM-s& 3 &3.42& 3.46 & 7.68  5.86 \\
& 5 & 3.24& 3.42  &7.56 5.83    \\
     \hline
     FM & 3 &3.38 & 3.42  &  7.59 5.79     \\
 & 5 &3.32 & 3.26 & 7.27 5.64   \\
   \hline
\end{tabular}
\end{center}
\end{table}

\begin{table}[h]\footnotesize
\caption{Ionic contribution $\varepsilon^p$ to static dielectric constant obtained by means of finite-difference approach for bulk systems.} \label{tab:ionic}
 \def\arraystretch{1.5}
\begin{center}
\begin{tabular}{  c|  c|  c }
Magn. & U [eV] &  $\varepsilon^p_{xx}$\:$\varepsilon^p_{yy}$\:$\varepsilon^p_{zz}$   \\
   \hline
AFM-N  & 3 & 4.93 4.84 0.46   \\
       & 5 & 4.60  4.52 0.46 \\
  \hline
AFM-z  & 3 &  5.03 4.69 0.45 \\
       & 5 & 4.64 4.44 0.46   \\
    \hline
AFM-s  & 3 & 5.04 5.16 0.46 \\
       & 5 &  4.66 4.67 0.46\\
     \hline
FM     & 3 & 5.14 5.05 0.45\\
       & 5 & 4.64 4.44 0.46   \\
   \hline
\end{tabular}
\end{center}
\end{table}

The 2D polarizability $\chi_{\parallel}$ is calculated as it was recently proposed in Ref. \cite{PhysRevB.88.045318}$ as \varepsilon (L_c)=1+\frac{4\pi \chi^{\parallel}}{L_c}$, where we neglect the higher order terms and $\varepsilon$ is the in-plane dielectric constant, whereas  $L_c$ is the interlayer separation between the centers of adjacent layers (in the case of monolayer L$_c$=c, where c is  lattice contant, in the case of bulk Lc is a vertical distance between the centers of adjacent layers). Note, that 2D polarizability calculated for ML and bulk give similar results (see Table \ref{tab:dielectric}), justifying this approach.

\subsubsection*{Selection rules}

The direct interband momentum (optical) matrix elements between the VBM and CBM at \textbf{k} point ($\mathbf{p}_{cv\mathbf{k}}$) are computed from the wave function derivatives using density functional perturbation theory \cite{PhysRevB.73.045112}. The dipole selection rules are 
$\left | \mathbf{x} \cdot \mathbf{p}_{cv\mathbf{k}}  \right |^2 >0 \leftrightarrow x$, $\left | \mathbf{y} \cdot \mathbf{p}_{cv\mathbf{k}}  \right |^2 >0 \leftrightarrow y$, $\left | \mathbf{z} \cdot \mathbf{p}_{cv\mathbf{k}}  \right |^2 >0 \leftrightarrow z$, with the linear polarization given by $\mathbf{x}=(1,0,0)$, $\mathbf{y}=(0,1,0)$ and $\mathbf{z}=(0,0,1)$ (for more details see \cite{PhysRevB.101.235408}).





\subsection{Exciton calculations}

The excitonic binding energies (shown in Fig.~2 of the main text) are obtained via the effective Bethe-Salpeter 
equation (BSE)\cite{Rohlfing2000PRB,Tedeschi2019PRB,FariaJunior2019PRB}. Considering excitons arising from one conduction and one valence band with  (anisotropic) parabolic dispersion, the BSE can be written as
\begin{equation}
\left[\Delta_{cv}(\vec{k})-\Omega_{N}\right]\!A_{N}(\vec{k})-\sum_{\vec{k}^{\prime}}\text{V}(\vec{k},\vec{k}^{\prime})\,A_{N}(\vec{k}^{\prime})=0 \, ,
\label{eq:BSE}
\end{equation}
in which $\Delta_{cv}(\vec{k}) = E_{c}(\vec{k})-E_{v}(\vec{k})$,  $E_{c}(\vec{k})$ and $E_{v}(\vec{k})$ are the conduction and valence 
bands that constitutes the exciton (in principle, from different points of the Brillouin zone), $\Omega_N$ and $A_{N}(\vec{k})$ are the 
energy and the envelope function of the $N$-th exciton state, respectively, 
and the electron-hole interaction is given by the potential 
$\text{V}(\vec{k},\vec{k}^{\prime})$. Although the BSE given in 
Eq.~(\ref{eq:BSE}) is general and applies to the monolayer and bulk cases, 
the energy bands and electron-hole interaction potential are different for each case. 
For example, in monolayers the wavevector $\vec{k}$ is a two-dimensional vector restricted 
to the plane and $E_{c}(\vec{k})$, $E_{v}(\vec{k})$ and 
$\text{V}(\vec{k},\vec{k}^{\prime})$ (the Rytova-Keldysh 
potential\cite{Rytova1967MUPB,Keldysh1979JETP}) take the specific form 

\begin{align}
E_{c}(\vec{k}) & =E_{c}+\frac{\hbar^{2}}{2m_{e}}\left(\frac{k_{x}^{2}}{m_{c,x}^{*}}+\frac{k_{y}^{2}}{m_{c,y}^{*}}\right) \, ,\nonumber \\
E_{v}(\vec{k}) & =\frac{\hbar^{2}}{2m_{e}}\left(\frac{k_{x}^{2}}{m_{v,x}^{*}}+\frac{k_{y}^{2}}{m_{v,y}^{*}}\right)  \, , \nonumber \\
\text{V}(\vec{k},\vec{k}^{\prime}) & =\frac{1}{\mathcal{A}}\frac{e^{2}}{2\varepsilon_{0}}\frac{1}{\varepsilon\left|\vec{k}-\vec{k}^{\prime}\right|+r_{0}\left|\vec{k}-\vec{k}^{\prime}\right|^{2}}  \, ,
\end{align}
in which $\hbar$ is the Planck's constant, $m_0$ is the free electron mass, $E_c$ is the band edge of the conduction band, $m^*_{c(v),x(y)}$ is the effective mass 
of conduction (valence) band along the $k_x$ ($k_y$) directions, $\mathcal{A}$ is the unit area, $e$ is the 
electron charge, $\varepsilon_{0}$ is the vacuum permittivity, $r_{0}$ is the 
screening length of the 2D material ($r_0 = 2\pi \chi^{\parallel}_{ML}$), and 
$\varepsilon$ is the effective dielectric constant that takes into account 
the dielectric screening of the surroundings ($\varepsilon=1$ is a monolayer in vacuum). For the bulk case, the wavevector $\vec{k}$ is a three-dimensional vector and $E_{c}(\vec{k})$, $E_{v}(\vec{k})$ and $\text{V}(\vec{k},\vec{k}^{\prime})$ (the anisotropic Coulomb potential\cite{Landau2013}) are then given by

\begin{align}
E_{c}(\vec{k}) & =E_{c}+\frac{\hbar^{2}}{2m_{0}}\left(\frac{k_{x}^{2}}{m_{c,x}^{*}}+\frac{k_{y}^{2}}{m_{c,y}^{*}}+\frac{k_{z}^{2}}{m_{c,z}^{*}}\right) \, , \nonumber \\
E_{v}(\vec{k}) & =\frac{\hbar^{2}}{2m_{0}}\left(\frac{k_{x}^{2}}{m_{v,x}^{*}}+\frac{k_{y}^{2}}{m_{v,y}^{*}}+\frac{k_{z}^{2}}{m_{v,z}^{*}}\right) \, ,  \nonumber \\
\text{V}(\vec{k},\vec{k}^{\prime}) & =\frac{1}{\mathcal{V}}\frac{e^{2}}{\varepsilon_{0}}\left[\sum_{a}^{\left\{ x,y,z\right\} }\varepsilon_{aa}\left(k_{a}-k_{a}^{\prime}\right)^{2}\right]^{-1}  \, , 
\end{align}
in which $\mathcal{V}$ is the unit volume, $\varepsilon_{aa}$ is the static dielectric constant along the $a$ direction ($a=\left\{ x,y,z\right\}$, and particularly, $\varepsilon_{xx}=\varepsilon_{yy}\neq\varepsilon_{zz}$).

We solved the BSE numerically considering a $k$-grid of $-k_{L}$
to $k_{L}$ in every dimension sampled with $(2N_{k}+1)^{n}$ with
$n=2$ for monolayers and $n=3$ for bulk. The final values of the exciton
binding energies are then obtained using a linear extrapolation of
the values calculated in the $k$-grid sampled with different number
of points. For instance, in the monolayer case we used $k_{L}=0.6\;\textrm{Å}^{-1}$
and $N_{k}=\left\{ 60,61\right\} $. In the bulk case, we used $k_{L}=0.5\;\textrm{Å}^{-1}$
and $N_{k}=\left\{ 19,20\right\} $. All the numerical 
inputs for the BSE, i.e., effective masses and dielectric constants are extracted from the first-principles calculations performed here. Particularly, the values 
for the effective masses in the monolayer case are taken from Table \ref{tab:ML}. The effective masses for the bulk case are taken from Table \ref{tab:Bulk} and rotated to the $k_x$, $k_y$ and $k_z$ using the angles, also given in the table. The values for the 2D polarizability of the monolayers, $\chi^{\parallel}_{ML}$, and the bulk static dielectric constants are given in Table~\ref{tab:dielectric}.


\section{Results}

\subsection{DFT studies.}
 \textit{The choice of the U parameter}. The standard exchange-correlation functionals such LDA, GGA are known to inadequately describe strongly  correlated systems which contain transition metals (3d states), as well as are known to underestimate the bands gaps of semiconductors. In order to correctly describe the $3d$ states of Mn ions, we used DFT+U formalism  proposed by Dudarev \cite{PhysRevB.57.1505}, where exchange coupling $J$ is considered via an effective Hubbard $U_{eff}$ coupling  constant $U_{eff}=U-J$, denoted throughout the paper as U. The DFT+U method is essentially empirical, in the sense that $U$ parameter must be provided. It is worth to mention that there is no standard procedure of obtaining U value for strongly correlated materials.
 
\begin{figure}[h]
\centering
\includegraphics[width=0.4\textwidth]{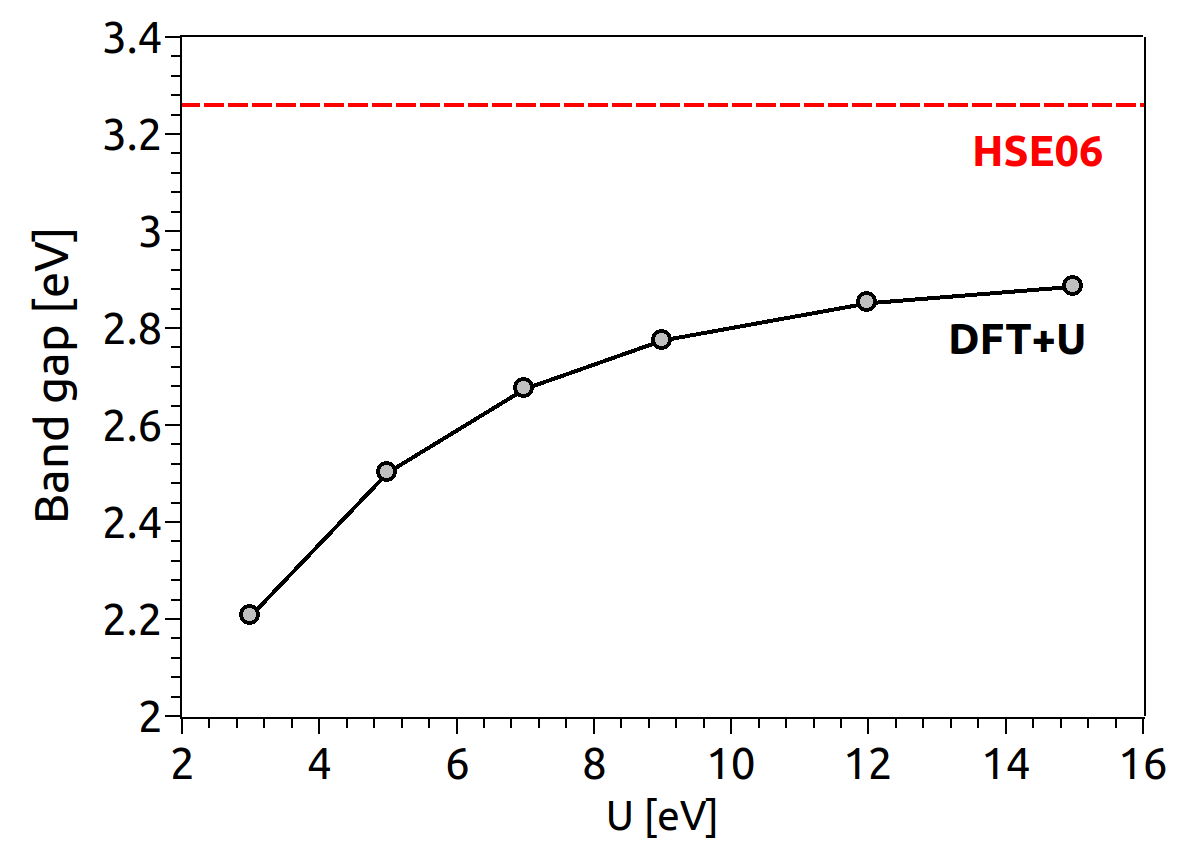}
\caption{\label{Ubands} Dependence of the band gap on the Hubbard U parameter obtained for the ground state of the monolayer. The dotted red line indicates the result obtained for the monolayer and using hybrid functional HSE06.}
\end{figure}

\begin{figure}[h]
\centering
\includegraphics[width=0.5\textwidth]{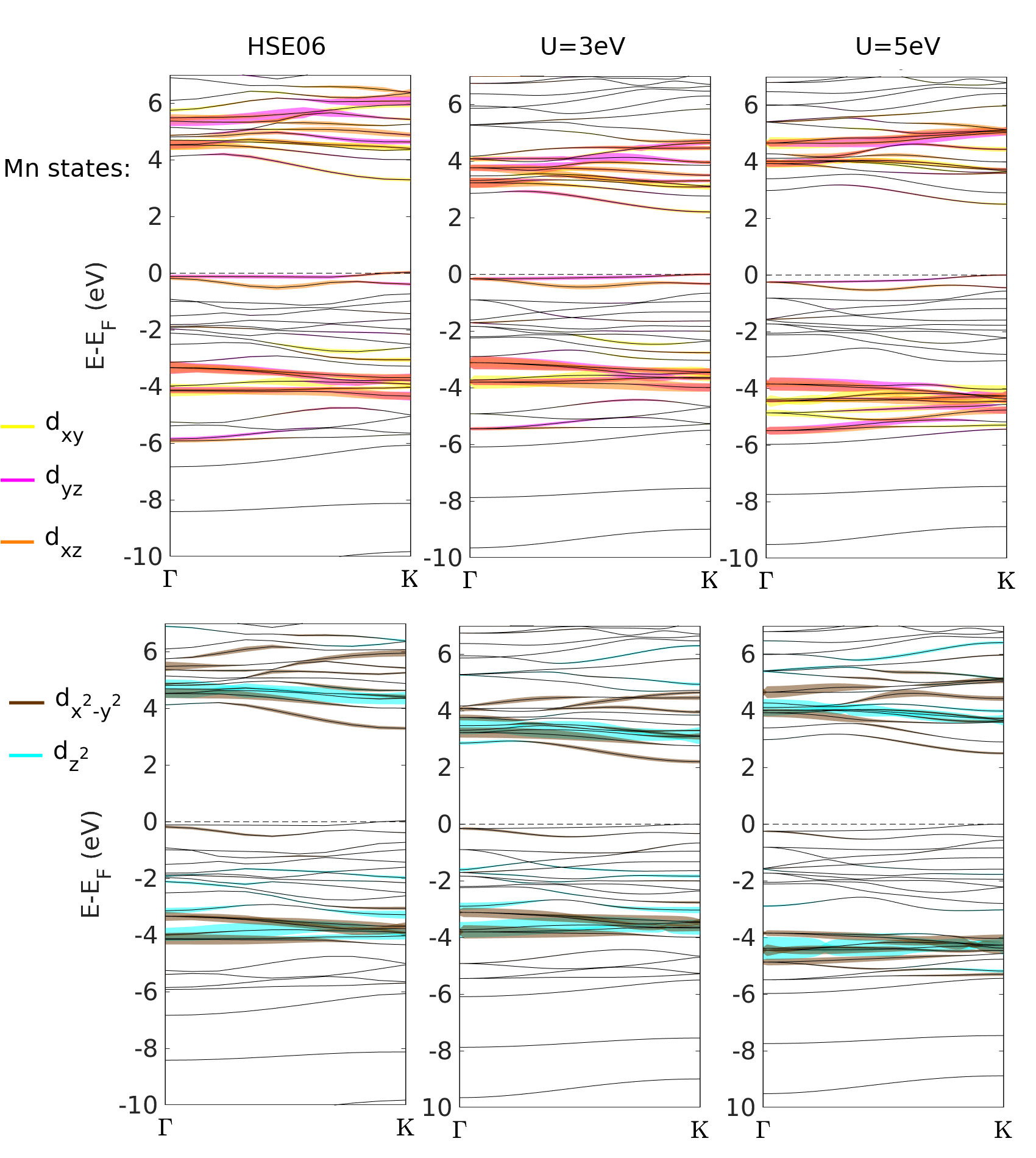}
\caption{\label{bandsProj} Bands projections of the 3d Mn states for: (A) HSE06 hybrid functional,(B) Hubbard U=3 eV, and (C) U=5 eV for the magnetic ground state of ML. }
\end{figure}

 Firstly, we present how the band gaps depend on the Hubbard U parameter, for the magnetic groundstate (AFM-N) (see Fig. \ref{Ubands}). Note, that the band gap increases as a function of U, approaching the maximal values of 2.9 eV, for nonphysically high value of U=10 eV. 
 Note, that it is not possible to chose the parameter U, for which the band gap would approach the experimental value for the bulk 3.0 eV \cite{Du2016}, or it would be equal to the band gap obtained from hybrid functional HSE06, which is equal to 3.26 eV for monolayer (similar value have been reported previously \cite{10.1039/C9RA09030D}). 
 In the case of the exciton calculations employed in this study, the crucial issue is the dispersion of the bands near the Fermi level. Thus, we present the comparison between the electronic structure obtained for using hybrid functional HSE06 and two different values of Hubbard U, namely, U=3 eV and U=5 eV (see Fig. \ref{bandsProj}). 
 Despite the fact, that the band gap in DFT+U approach is underestimated, however, the dispersion of the bands looks similar, especially for the case of U=3 eV, in comparison to HSE06 functional, similar results have been reported recently in \cite{Yang2020RSC}. Thus, the U=3 eV have been chosen for the further calculations, and also U=5 eV for the comparison purposes and to assess the impact of U parameter on physical properties.
 
\textit{Electronic properties}. The summarized results of electronic properties for monolayers and bulk systems in respect to employed here magnetic states are collected in the Table S4 and Table S5, respectively. In order to determine the VBM and CBM of the bulk systems, we precisely probe the half of an entire first BZ zone, to directly show from which k-points the transition occurs. The exact coordinates of the valence band maximum and conduction band minimum are listed in Table S5. 
Our results reveal (see Table S5) that the bulk MnPS$_3$ is a wide semiconductor \cite{Brec1979, Du2016, PhysRevB.94.184428}, and the predicted band gap for the groundstate (AFM-N) is consistent with previously reported values for the bulk \cite{10.1039/C9RA09030D}. Moreover, the band gaps depend on the magnetic state of the Mn ions (see Tables S4 and S5), namely, differ from few tens to few hundredths of electronovolts from each other, independently of U parameter.  The smallest band gap is obtained in case of AFM-s spin configuration.
In addition, the  band gaps of bulk are smaller than monolayer ones approximately hundredths of meV, which is consistent with band gap trend of the other layered materials. The similar dependence of Hubbard U  on the band gaps are observed for the monolayer and bulk systems.

\begin{table*}[h]
\label{tab:ML}
\caption{ Calculated band gaps, valence and conduction edges, and effective masses of monolayer for various magnetic states of Mn ions. The symbol D, sD, ID in parenthesis indicate the direct band gap, semi direct band gap, indirect band gap, respectively.  Note that the sD band gap occurs in  very close vicinity of high symmetry K point  in BZ, and in good approximation can be further considered as direct. The $m_1^{\parallel}$ and $m_2^{\parallel}$ are determined along  (1 0 0) and (0 1 0) directions in Cartesian reciprocal space, respectively and are given in $m_e$ unit.}
  \def\arraystretch{1.5}
 \begin{center}
  \begin{tabular}{  c |c|c|c|c| c|c}
    \hline
 magn.  & U [eV]& band gap [eV]  & VBM & CBM &  $m_{1}^{\parallel}$, $m_{2}^{\parallel}$ - hole [$m_e$]&   $m_{1}^{\parallel}$, $m_{2}^{\parallel}$ - electron [$m_e$]\\
    \hline
AFM-N & 3 & 2.206 (D) &K&K &-1.26, -1.25 & 0.68, 0.68\\
& 5 &2.502 (D)  &K&K & -0.96, -0.95 &  0.58, 0.58 \\
\hline
 AFM-z & 3 & 1.969 (sD)  &near K&near K & -1.34,  -0.90 & 0.61 0.90 \\
 & 5 & 2.345 (sD)&near K & near K & -1.03, -0.85&   0.55,  0.71 \\
 \hline
 AFM-s & 3 & 1.795 (D) &K&K & -0.79,  -0.77& 0.77, 0.53 \\
 & 5 &  2.207 (D) &K&K &-0.78,  -0.73&  0.62, 0.51 \\
 \hline
 FM ($\uparrow$) & 3 & 2.268 (D) &K&K &-0.60, -0.60 & 0.40, 0.40 \\
 & 5 & 2.412 (ID)  &K &$\Gamma$&-0.60, -0.60&   0.30, 0.30\\
  \hline
\end{tabular}
\end{center}
\end{table*}

In the case of the effective mass calculations, it is worth to mention, that for the monolayer systems, the principle reciprocal axes coincide with the Cartesian reciprocal axes, and the $m_1^*$ and $m_2^*$ components can be considered as in-plane (001) components (within the layer) denoted here as $m_1^{\parallel}$ and $m_2^{\parallel}$, respectively. Note, that $m_3^*$ is out-of-plane (001) component (out of the layer) denoted as $m_3^{\perp}$ and for all monolayer systems approaches infinity and thus, it not presented. However,  for the bulk case, only $m_2^*$ can be considered as in-plane (001) component ($m_2^{\parallel}$), the other two $m_1^*$, $m_3^*$ are rotated counterclockwise at the angle of $\Theta$ from [100] and [010] crystallographic directions in reciprocal space, respectively (rotation about y-axis, $\Theta$ angles are listed in Table S5).
In addition, the AFM-z and AFM-s magnetic configurations exhibit the anistropic behaviour of in-plane effective masses. For  AFM-s and AFM-z cases, the in-plane [100] direction can be considered as lees mobile carrier direction (heavier carriers), than for [010] crystallographic direction, indicating the differences of the conductivity that could be measured within the monolayer plane. Note, that the in-plane components of effective masses can reveal the structural symmetry of MnPS$_3$, namely the $m_1^{\parallel}=m_1^{\parallel}$ can indicate the presence of hexagonal symmetry for FM, AFM-N cases, whereas the $m_1^{\parallel}\neq m_1^{\parallel}$  breaking of this symmetry observed in AFM-s, AFM-z cases.

\begin{table*}[h]
\small
\label{tab:Bulk}
\caption{Calculated band gaps for various magnetic states of Mn atoms for the bulk. The fractional coordinates of the valence band maximum (VBM) and conduction band minimum (CBM) given in reciprocal lattice vectors (RLV) are collected in the columns 5 and 6, respectively. Note, that the coordinates of the  high symmetry K point folded to the "e" supercel (sc) for AFM-z and AFM-s is (-1/3 0 0.110) in RLV, whereas in the case of"p" supercell for FM and AFM-N states is (1/3,-1/3,0.22) in RLV, which have been calculated accordingly with the values given in Table \ref{tab:Kpoints}. The direction of the principal axes $m^*_1$ and $m^*_3$ are counterclockwise rotated at the angle $\Theta$ from [100] and [001]   directions, respectively. }
  \def\arraystretch{1.5}
 \begin{center}
  \begin{tabular}{  c |c| c|  c| c|c|c|c|c|c}
    \hline
 magn. state & U [eV] & band gap [eV]  & k-point of VBM & k-point of CBM  & sc &mass&hole [$m_e$] &electron  [$m_e$]&\textbf{electron}  [$m_e$]  \\
 \hline
AFM-N & 3 &2.041 (ID) (\textbf{2.106 D})  &(1/3 -1/3 0.264) & (-1/3 1/3 0.426) & p &$m_1$ &-1.15 &0.60  &\textbf{0.80 }\\
&&$\Delta_{gap}^{ID-D}=65$ meV&&&&$m_2^{\parallel}$&-1.27 & 0.69 &\textbf{0.70 }\\
&&&&&&$m_3$ & -2.25 & 5&\textbf{-7.27 }\\
\cline{7-10}
&&&&&&$\Theta$ & 57.0 &  2.9& \textbf{5.58}\\
\cline{2-10}

& 5 &2.352 (ID) (\textbf{2.369 D})& (1/3 -1/3 0.279)& (-1/3, 1/3, 0.44) & p &$m_1$ & -1.15 &0.51 &\textbf{0.70}\\
&&$\Delta_{gap}^{ID-D}=17$ meV&&&&$m_2^{\parallel}$&-0.96 & 0.59&\textbf{0.60}\\
&&&&&&$m_3$ & -2.11&4.17&\textbf{-6.04 }\\
\cline{7-10}
&&&&&&$\Theta$ & 30.0 & 2.8 &\textbf{5.01}\\
\hline

 AFM-z &  3 &1.828 (ID) (\textbf{1.840 D})&(-0.300 0 0.289)  & (1/3 0 0.441) &e &$m_1$ & -1 &0.56 &\textbf{0.62 }\\
 &&$\Delta_{gap}^{ID-D}=12$ meV&&&&$m_2^{\parallel}$ &-0.86  & 0.98 &\textbf{0.99 }\\
&&&&&&$m_3$ &-4.13& 5.09&\textbf{-31.16}\\
\cline{7-10}
&&&&&&$\Theta$  & 30.9 & 2.7& \textbf{5.85}\\
\cline{2-10}
 
 & 5 & 2.212 (ID) (\textbf{2.225 D})&  (-1/3 0 0.317) & (1/3 0 0.454) & e&$m_1$&-0.86  & 0.49 &\textbf{0.59}\\
  &&$\Delta_{gap}^{ID-D}=13$ meV&&&&$m_2^{\parallel}$ & -0.91& 0.74 &\textbf{0.73}\\
&&&&&&$m_3$ &-3.63 & 4.01 &\textbf{107.52 }\\
\cline{7-10}
&&&&&&$\Theta$  & 16.2 & 3.0& \textbf{5.01}\\
 \hline
 
 AFM-s& 3 &1.653 (ID) (\textbf{1.656 D})& (0.367 0 0.427) & (1/3 0 0.482) &e &$m_1
$&-0.63  &0.7 &\textbf{0.76 }\\
&&$\Delta_{gap}^{ID-D}=3$ meV&&&&$m_2^{\parallel}$ &-0.87  &0.55&\textbf{0.52 }\\
&&&&&&$m_3$ & -1.97&5.29  &\textbf{5.67 }\\
\cline{7-10}
&&&&&&$\Theta$   & 2.6 &  6.5& \textbf{4.44}\\
 \hline
 
 & 5 &2.066 (ID) (\textbf{2.067 D})&(0.367 0 0.441) &(1/3 0 0.482) & e&$m_1
$&-0.64 & 0.56 &\textbf{0.60} \\
&&$\Delta_{gap}^{ID-D}=1$ meV&&&&$m_2^{\parallel}$ &-0.92 & 0.53  &\textbf{0.52}\\
&&&&&&$m_3$ &  -2.17 )&4.12&\textbf{4.35}\\
\cline{7-10}
&&&&&&$\Theta$   & 4.4  &4.9& \textbf{3.63}\\
 \hline
 FM ($\uparrow$) & 3 &2.096  &(-1/3 0 0.414)  & (1/3 0 0.414)& e &$m_1$ &-0.54  & 0.44 &\textbf{0.43 }\\
&&&&&&$m_2^{\parallel}$ &-0.56  &0.42 & \textbf{0.44 }\\
&&&&&&$m_3$ & -4.05 & -4.20 &\textbf{-1.88 }\\
\cline{7-10}
&&&&&&$\Theta$    &   5.3 &  0& \textbf{3.14}\\
\cline{2-10}
 & 5 & 2.276 (ID) (\textbf{2.419 D}) &(-1/3 0 0.400)& (0 0 0) &  e&$m_1$&-0.54  &0.26 &\textbf{0.43 }\\
 && $\Delta_{gap}^{ID-D}=143$ meV&&&&$m_2^{\parallel}$ &-0.57  &0.27  &\textbf{0.45 }\\
&&&&&&$m_3$ &-4.15 &1.63 &\textbf{-1.94 }\\
\cline{7-10}
&&&&&&$\Theta$& 5.2 & 4.3 &\textbf{3.4}\\
\hline
\end{tabular}
\end{center}
\end{table*}

\subsection{Exciton binding energies}

Here we provide the exciton binding energies for bulk (Table \ref{tab:exc_bulk}) and monolayer (Table \ref{tab:exc_ML}) of MnPS$_3$. 
Although, we provide the exciton binding energy as a positive value, it lies below the single-particle band gap.

\begin{table}[h]
\caption{Exciton binding energies for bulk MnPS$_3$.}
\label{tab:exc_bulk}
\begin{ruledtabular}
\begin{tabular}{ccccc}
 & \multicolumn{2}{c}{direct} & \multicolumn{2}{c}{indirect}\tabularnewline
 & U=3eV & U=5eV & U=3eV & U=5eV\tabularnewline
 \hline
AFM-N & 177.09 & 185.21 & 150.83 & 151.31\tabularnewline
AFM-s & 137.15 & 135.87 & 136.15 & 133.97\tabularnewline
AFM-z & 177.33 & 170.77 & 159.94 & 150.26\tabularnewline
FM    &        &        & 160.69 &  92.43\tabularnewline
\end{tabular}
\end{ruledtabular}
\end{table}
\begin{table*}[h]
\caption{Exciton binding energies for monolayer MnPS$_3$ considering different values of the effective dielectric constant of the surroundings.}
\label{tab:exc_ML}
\begin{ruledtabular}
\begin{tabular}{ccccccccc}
 & \multicolumn{2}{c}{AFM-N} & \multicolumn{2}{c}{AFM-z} & \multicolumn{2}{c}{AFM-s} & \multicolumn{2}{c}{FM}\tabularnewline
$\varepsilon$ & U=3eV & U=5eV & U=3eV & U=5eV & U=3eV & U=5eV & U=3eV & U=5eV\tabularnewline
\hline
1 & 1027.01 & 1017.06 &    1022.86 & 1025.83 &   971.85 & 987.48 &   881.07 & 843.90\tabularnewline
2 &  672.11 &  651.27 &     668.75 &  658.32 &   623.32 & 625.72 &   545.45 & 512.12\tabularnewline
3 &  493.07 &  469.35 &     490.12 &  475.83 &   450.17 & 447.55 &   383.02 & 354.16\tabularnewline
4 &  382.86 &  359.18 &     380.00 &  364.55 &   344.89 & 339.84 &   286.87 & 262.22\tabularnewline
5 &  307.39 &  285.29 &     304.96 &  290.02 &   273.68 & 268.17 &   223.63 & 202.21\tabularnewline
\end{tabular}
\end{ruledtabular}
\end{table*}

\newpage
\newpage
\bibliography{MnPS3}

\end{document}